\documentclass[a4paper,onecolumn,showpacs,showkeys,amsmath,amssymb,11pt]{revtex4}
\usepackage{graphicx}% Include figure files
\usepackage{epsfig}
\usepackage{dcolumn}% Align table columns on decimal point
\usepackage{bm}% bold math
\usepackage{color}
\usepackage{ifpdf}
\usepackage{hyperref}

\begin{document} 
\title{LISACode: A scientific simulator of LISA} 
\author{Antoine Petiteau, G\'erard Auger, Hubert Halloin, Olivier Jeannin, Eric Plagnol}
\affiliation{APC, UMR 7164, Universit\'e Paris 7 Denis Diderot\\10, rue Alice Domon et L\'eonie Duquet, 75025 Paris Cedex 13, France}
\email{antoine.petiteau@apc.univ-paris7.fr}
\author{Sophie Pireaux, Tania Regimbau, Jean-Yves Vinet}
\affiliation{ARTEMIS, Observatoire de la C\^{o}te d'Azur-C.N.R.S., 06304 Nice, France}
\today

% *********** 
% *Abstract * 
% *********** 
\begin {abstract}  
A new LISA simulator (LISACode) is presented. Its ambition is to achieve a new degree of sophistication allowing to map, as closely as possible, the impact of the different sub-systems on the measurements. LISACode is not a detailed simulator at the engineering level but rather a tool whose purpose is to bridge the gap between the basic principles of LISA and a future, sophisticated end-to-end simulator. This is achieved by introducing, in a realistic manner, most of the ingredients that will influence LISA's sensitivity as well as the application of TDI combinations. Many user-defined parameters allow the code to study different configurations of LISA thus helping to finalize the definition of the detector. Another important use of LISACode is in  generating time series for data analysis developments.
\end{abstract}

\maketitle 

%\tableofcontents 

%\indent 
% **************** 
% * Introduction * 
% **************** 
\section{Introduction} 
\label{SIntro} 

The LISA mission aims at detecting gravitational waves (GW) from space, using 3 detectors  
located on the summits of a triangle whose center orbits around the Sun,  
following the Earth, at a distance of $\sim 50\times10^{6}$ km. The distance between 
spacecraft is of the order of $5\times10^{6}$ km. Located inside each spacecraft, two (or one of two) proof masses are kept on a geodesic orbit by a drag-free system (DFS): a system that compensates, through the use of micro-thrusters, for the external forces acting on the spacecraft by keeping it at its proper position with respect to the proof masses.

The triangular configuration, which depends on the orbits each spacecraft follows, is not rigid and will both rotate around its center and ``flex'' (i.e. the distance between  spacecraft is time dependent) with a period of one year. On a much smaller length scale,  
picometers, and at frequencies between $10^{-4}$ and $10^{-1}$ Hz (the current instrumental 
bandwidth), the distance between spacecraft will be modified by the passage of a gravitational wave (GW). The detection of such GW will be achieved by interferometric measurements giving the phase (frequency) difference between local and distant (i.e. coming from another spacecraft) laser beams on each spacecraft. There are therefore, 6 independent laser beams and 12 interferometric 
measurements: 6 between local lasers and 6 between local and distant lasers.   
A detailed description of LISA, of its principles and of the different noise sources,  
is given in the LISA Pre-Phase A Report \cite{PrePhaseAReport}.

The LISA detector system is complex and its sensitivity depends to a large extent on the  
different noise contributions coming from the DFS and the interferometric measurements. For these reasons and because a laboratory replica of the system is not totally achievable, the exact evaluation of the performances of LISA can only be studied with computer simulations of the different processes involved. Two such simulators have been elaborated in the US: Synthetic LISA \cite{SyntheticLISA} and  LISA Simulator \cite{LISASimulator}.

The present status of the LISA project warrants that precise evaluations of its performances be performed, compared and checked through different simulation softwares. The software implementation of the LISA method and, for example, the introduction of the noise, the measurement of the phase, the proper time stamping of the measurements and the reduction of the laser phase noise by the TDI method are all non trivial and may depend on the manner in which different "technologies", not always perfectly defined at the present status of the project, are implemented at the software level. Of the two simulators mentioned above, Synthetic LISA\cite{SyntheticLISA} is the one that appears closest to mapping these technologies. However it remains necessary to develop similar software simulators in order to push this implementation further and to check that the quality and the sensitivity of the LISA detector is independent of the structure and numerics of these codes.

This paper presents such a new LISA simulator, LISACode, whose ambition is twofold: i) a modular,  
semi-technical, implementation of the different ingredients, and particularly the different  
noise sources and the phasemeter, which will affect the quality of the measurements and ii) the application of the TDI (Time Delay Interferometry), on the phasemeter outputs, which allows to extract the GW signal from the overwhelming laser noise. The main inputs of the code are user-defined or built-in GW sources and the various parameters which define the noise levels and the orbits of the spacecraft. The main outputs of LISACode are time sequences, with a user-defined sampling step, of the basic (phasemeter) measurements and/or time sequences of  user-defined TDI generators.

After this introduction, the second section of this paper presents the structure and different elements of LISACode. The third section gives some examples of studies that LISACode can be used for and a last section presents a number of GW pre-defined in LISACode. Two appendices describe the reference frames used in LISACode and the CPU performance of the code in various configurations. The LISACode software package and a more detailed description, can be downloaded from the LISA-France web site at http://www.apc.univ-paris7.fr/LISA-France/analyse.phtml. 

% ************************************ 
% * Basics principles of LISACode * 
% ************************************ 
\section{LISACode} 
\label{SPrincip} 
% 
% General 
% 
%\subsection{General} 
%\label{SSPrincipGeneral} 
LISACode is not a detailed simulator at the engineering level but rather a tool  
whose purpose is to bridge the gap between the basic principles of LISA and a, future,  
sophisticated end-to-end simulator at mission level. This is achieved by introducing, in a realistic manner, most of the ingredients that will influence its sensitivity. The end product of LISACode are the time series of the phasemeter outputs and their fluctuations arising both from the instrumental noises and from the statistical fluctuations inherent to any measurement process.

% 
% The Structure of LISACode
% 
\subsection{The Structure of LISACode} 
\label{structure} 

\begin{figure}[!ht] 
\centering \includegraphics[width=8.6cm]{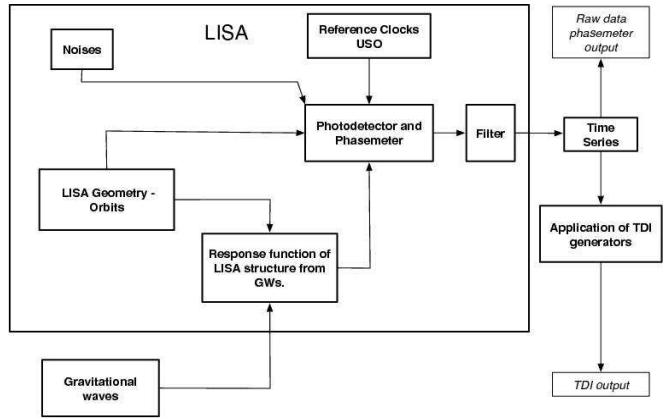} 
\caption{\small Schematic diagram of the LISACode software simulator.} 
\label{Structure} 
\end{figure}

The schematic structure of LISACode is presented in figure \ref{Structure}. The core of the program which simulates LISA is shown inside the box.  The GW calculations and the application of TDI, although they are part of the code, are not part of LISA and can be considered as external modules. The basic output of the code are time series of the phasemeters. These time series can be calculated with a variety of different noise inputs with or without the presence of a GW. These time series can be post-treated by a number of pre-defined or user-defined TDI combinations. At the end of this section, various sensitivity curves for different TDI combinations will be presented showing the numerical accuracy that can be obtained by LISACode.
	
\

As stated above, the aim of LISACode is to follow, as closely as possible, the technological design of the future detector. There are, at least, two ingredients that are fundamental in this respect : the different noises that intervene and limit the sensitivity of LISA and the precision of the measurement of the phase (or frequency) by the $phasemeter$ module. Associated to this are the $USO$ (Ultra Stable Oscillator, the reference clock on each satellite) which will impact both on the time-stamping of the time series and on the precision of the phasemeter. We discuss below some of the more important features of LISACode.

\subsection{The Orbits and photon flight paths} 
\label{orbits} 
The orbits of the three spacecraft are calculated using the analytical formulation 
of  \cite{OrbitLISA}.  
These orbits include both the rotation of the LISA constellation and the flexing of the  
distance between the spacecraft. At the origin of time ($t=0$), the center of gravity of the LISA constellation points towards the vernal point. The photon flightpath between two satellites can be computed at several orders in an expansion 
in powers of the  Sun's gravitational potential, giving increasing degrees of accuracy \cite{OrbitRGOpticalLink}.   
At the lowest order, the light propagation time is computed using the distance  
between the spacecraft. Further precision can be obtained by taking into account the  
Sagnac effect, the aberration and  other General  Relativity effects\cite{OrbitRGOpticalLink}.  

\subsection{The Noises} 
\label{noises} 

In LISA, the noises can be divided in two groups depending on wether they are transported from one satellite to another (allowing their suppression, to a certain level, by the application of TDI) or wether they impact locally (and cannot be suppressed).\\

\begin{itemize}
\item
  \textbf{External noises}: laser noises that are transported from one satellite  
to another and which will be suppressed by TDI. At this moment in time, the laser noise is considered to be a bandwidth limited white noise whose power spectral density is $30 {\rm Hz/\sqrt {Hz}}$ \cite{30hz}. It is almost certain that the effective laser noise will present a more complicated frequency dependence, notably at low frequencies. This will be implemented as soon as more information on this subject is available.
\item
  \textbf{Internal noises}: inertial mass, Ultra Stable Oscillators (USO), 
  shot and optical path noise. These will be transmitted through the phasemeter transfer function and will impact on the precision of the measurements.
\end{itemize}

Table \ref{table_error} gives the different inputs to the internal noise for a phasemeter. The values of the second column are those given in reference \cite{PrePhaseAReport}. The third column indicates the noise used, by default, in LISACode. These values can be modified, for each individual phasemeter, by the user. Note that the noise in  
lines 3 to 8 are, in LISACode, summed up quadratically, for each phasemeter, into a unique term. The last column of table \ref{table_error} gives the assumed dependency of the different noise with the laser power and the distance between satellites. The $residual$ $laser$ $noise$ does not appear in LISACode as this was assumed to represent the residual noise after suppression of the laser noise by TDI \cite{LISAASD} which is treated explicitly by LISACode (see section \ref{tdi}).

When translated  to fractional frequency fluctuation ($\Delta \nu / \nu$ unit) all but the last noise given in this table have an $f$ dependence at the phasemeter level. In the same units, the inertial mass noise has an $1/f$ dependence.

A number of remarks have to be made on this table. Since it was issued, ESA \cite{LISAASD} has made a more refined study of the allowed noises and has subdivided them into different components. At the phasemeter level, all the $measurement$ noises (using the terminology of \cite{LISAASD}) have a $\sqrt{f^2+ a/f^2}$ frequency dependence, whereas the $acceleration$ noise has a $\sqrt{b/f^2+f^2/c+ a/f^6}$ frequency dependence. This subdivision of different noises and their frequency dependence are being implemented in LISACode but the results presented in this paper maintain the definition and frequency dependence given in table \ref{table_error}.

\begin{table}[htdp] 
\caption{Noise allocation budget based on Table 4.1 of  \cite{PrePhaseAReport}.  The second column gives errors listed in the PrePhaseAReport\cite{PrePhaseAReport}. The third column gives values used in LISACode in the same unit as the second column. The fourth column gives the default errors included in LISACode and their frequency dependence. These values can be  modified by the user.} 
\begin{center} 
\begin{tabular}{|c|c|c|c|c|c|} 
\hline 
\multicolumn{2}{|c|}{Error source}&Error&LISACode& LISACode input ($\delta \nu \over \nu $ unit) \\ 
\hline 
\multicolumn{5}{|c|}{Measurement Noise}  \\ 
\hline 
 & Detector shot noise&$11\times 10^{-12} {\rm m.{Hz}^{-{1 \over 2}}}$&$11$ & $2.3 \times 10^{-19}\left(f\over 1 Hz \right) \left(L \over 5 \times 10^{9}{\rm m} \right) \sqrt{1 {\rm  W}\over P }.{Hz}^{-{1 \over 2}}$\\ 
\cline{2-5}
 & USO & $5 \times10^{-12} {\rm m.{Hz}^{-{1 \over 2}}}$& &  \\  
\cline{2-3} 
 & Laser beam-pointing &$10 \times10^{-12} {\rm m.{Hz}^{-{1 \over 2}}}$& & \\ 
 & instability& & & \\ 
\cline{2-3} 
$s_{i}^{Meas. noise}$ & Laser phase measurement &$5\times10^{-12} {\rm m.{Hz}^{-{1 \over 2}}}$& $15.7$ &  $3.32 \times 10^{-19} \left(f\over 1 Hz \right).{Hz}^{-{1 \over 2}}$\\ 
& and offset lock& & & \\ 
\cline{2-3} 
& Scattered-light effects&$5\times10^{-12} {\rm m.{Hz}^{-{1 \over 2}}}$& &  \\ 
\cline{2-3} 
& Other substantial effects&$8.5\times10^{-12} {\rm m.{Hz}^{-{1 \over 2}}}$& & \\ 
\hline 
& Residual laser phase noise&$5 \times 10^{-12} {\rm m.{Hz}^{-{1 \over 2}}}$& not incl. &  \\ 
\hline 
\multicolumn{5}{|c|}{Acceleration Noise}  \\ 
\hline 
$\delta_i^{Acc. noise} $& Inertial Mass noise & $3 \times10^{-15}{\rm m.s^{-2}.{Hz}^{-{1 \over 2}}}$& $3 $ & $1.59 \times 10^{-24} \left( 1 Hz \over f\right).{Hz}^{-{1 \over 2}}$\\ 
\hline 
\end{tabular} 
\end{center} 
%\label{default} 
\label{table_error}   
\end{table}% 

\subsection{The Phasemeter} 
\label{phasemeter} 

\begin{figure}[!ht] 
\centering \includegraphics[width=10cm]{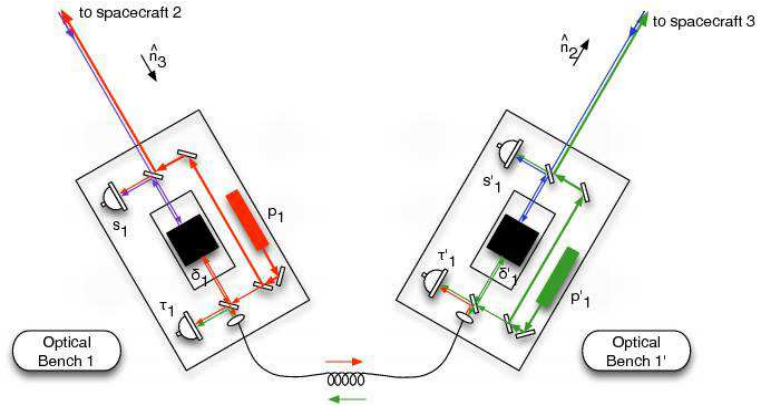} 
\caption{\small Schematic representation of the two optical benches of spacecraft 1 with the phasemeters ($s$ and $\tau$) the lasers $p$ and the inertial masses $\delta$ (from \cite{TDITinto}), see text (Color on line).}   
\label{SchBOSC1} 
\end{figure}

In LISA, measurements are performed at the phasemeter level (see fig. \ref{Structure}).  
A phasemeter sums up the effects of the gravitational strain and of the various noises.  
There are two phasemeters for each optical bench (see fig. \ref{SchBOSC1}): one ($s_i$) measures the phase between the external laser and the local one, the other ($\tau_i$) measures the phase between the local laser and the local laser of the other optical bench in the same spacecraft. For example, Equations \ref{s1LISACode},  
\ref{tau1LISACode}, \ref{s1pLISACode} and \ref{tau1pLISACode} give the structure of the signals  for the four phasemeters of spacecraft 1: 

\begin{eqnarray} 
s_1 &= &s_1^{GW}+ s_1^{Meas. noise}+D_3p_2^{'laser. noise}-p_1^{laser. noise} -2\delta_1^{Acc. noise}
\label{s1LISACode}\\ 
\tau_1 &= &p_1^{'laser. noise} -p_1^{laser. noise}- 2\delta_1^{'Acc. noise} 
\label{tau1LISACode}\\ 
s'_1 &= & {{s'}_1}^{GW}+ {{s'}_1}^{Meas. noise} + D'_2p_3^{laser. noise}-p_1^{'laser. noise} + 2\delta_1^{'Acc. noise}
\label{s1pLISACode}\\ 
\tau'_1 &= &p_1^{laser. noise} -p_1^{'laser. noise} + 2\delta_1^{Acc. noise} 
\label{tau1pLISACode} 
\end{eqnarray} 
where the unprimed (resp. primed) quantities refer to the link i$\rightarrow$ i+1 
(resp.  i$\rightarrow$ i-1)(see figure \ref{SchBOSC1}). For any function of time $f$, 
$D_{i}f = f(t - t_i)$, $D'_{i}f = f(t - t'_i)$, are the time delay operators, and 
the delays $t_i,t'_i$, the photon flight times in the two possible directions. 
$D_{i}$ takes into consideration the travel time in the direction $3 \rightarrow 2 \rightarrow 1$ and  $D'_{i}$ in the direction  $1 \rightarrow 2 \rightarrow 3$. $s_{i}^{GW}$ and ${s'}_{i}^{GW}$ are the gravitational signals on the laser link respectively from the spacecraft $i+1$ to the spacecraft $i$ and from the spacecraft $i-1$ to the spacecraft $i$. $s_{i}^{Meas. noise}$ groups the shot noise and other optical path noises (see table \ref{table_error}) received on phasemeter $s_{i}$. $p_{i}^{laser. noise}$ are the laser  noise and  $\delta_i^{Acc. noise}$ is the inertial mass noise (see table \ref{table_error}). All these signals are measured in relative frequency fluctuation unit ($\Delta \nu / \nu$ unit). 

The impact of the phasemeter measurements and errors on the final quality of the time series is a sensitive issue and it is a specific feature of LISACode that it attempts to tackle this technological problem as precisely as possible in view of our present knowledge the phasemeter. Traditionally, the phasemeter outputs will be transmitted to earth at a rate which may vary from $1 Hz$ to $3 Hz$. In order to eliminate the high frequency component (i.e. frequencies above 0.5 the transmitted data rate) the phasemeter \emph{must} be associated with a filter. In LISACode this filter is implemented in the following way: once the sampling rate of the phasemeter output is chosen by the user, the time series are calculated at a multiple (user-defined) of this frequency and a filter is applied to suppress the frequencies above a tenth of the sampling rate. The outgoing signal will therefore still have a (small) residual high frequency component (i.e. above the cut-off frequency) which will impact on, for example, the application of TDI. Although the final filtering properties associated to the phasemeter may depart from this implementation, it is considered that such features should be present in this kind of simulation in order to evaluate the impact it may have on the final performance of the detector and of the analysis. Figure  \ref{PSDXGWvsNoise} of section \ref{tdi} shows (red curve above) an example of the phasemeter+filter output for a sampling rate of $1 Hz$ and illustrates the cut-off obtained above $0.1 Hz$ .

\subsection{The Reference Clocks} 
\label{USO} 

The USO (or reference clock) noise impacts the measurements at two levels. As the USO (one in each satellite) will be used for the phasemeter module by giving it a reference signal, its noise will impact on the phasemeter output. This is accounted for by the corresponding noise given on the fourth line of Table \ref{table_error}. The USO is also used to $time-stamp$ the time series and will thus introduce an error in the application of TDI. In LISACode, the time stamping is affected both by this noise as well as by a, user-defined, offset and a drift with respect to barycentric time. At the phasemeter level, only the USO noise is considered. The more complicated effect of a drift is under study.

% 
% Time Delay Interferometry 
% 
\subsection{TDI and the LISA Sensitivity Curve} 
\label{tdi} 
Due to the large distance between the spacecraft, the interferometric measurements are performed via independent (local and distant) lasers. The laser noise level is typically of the order of $10^{-13}$ (in $\Delta \nu / \nu$ unit),  
whereas, the gravitational wave signal is closer to $10^{-21}$. This implies that the laser  
noise must be suppressed by more than 8 orders of magnitude. Today, this is achieved by a  
numerical method called \textbf{Time Delay Interferometry (TDI)}.  An extensive description  
of the TDI method, and of the different combinations, can be  
found for example in  \cite{TDITinto,TDIVinet,TDIGeometric} and \cite{TDIMoving}. TDI is fully implemented in LISACode.

The general philosophy of TDI is to linearly combine the phasemeter measurements  
obtained at different times and spacecraft in such a way as to eliminate numerically the  
laser noise. There are various linear combinations that allow for this. As an example the  
Michelson $X$ ($1^{st}$ generation) generator corresponds to the combination: 

\begin{equation} 
\begin{array}{lll} 
X &=& - s_1 - D_3 \ {s'}_2 - D_3 D_{3'} \ {s'}_1 - D_3 D_{3'} D_{2'} \ s_3  \\ 
& &+ {s'}_1 + D_{2'} \ s_3 + D_{2'} D_2 \ s_1 + D_{2'} D_2 D_3 \ s_{2'}
\end{array} 
\label{TDIGeneratorX} 
\end{equation}  
where $s_i$ is a phasemeter output (see eq. \ref{s1LISACode}) and $D_{i}$ is a delay operator based on the time of flight of the the laser over the link $L_i$:  
\begin{equation} 
D_{i} x(t) = x(t - L_i/c) 
\label{eq_delay}
\end{equation} 

Such combinations will tend to zero with a residual value which will depend on the level of the experimental noises (see section \ref{noises}) and on the accuracy with which the phasemeter time series can be interpolated at the corresponding delay times.

LISACode includes a number of predefined TDI combinations, for a fixed ($1^{st}$ generation) and a moving ($2^{nd}$ generation) LISA: $X$, $Y$ ,$Z$ , $\alpha$,  $\beta$, $\gamma$ and $\zeta$  as well as Relay, Monitor and Beacon.

\begin{figure}[!ht] 
\centering 
\includegraphics[width=8.6cm]{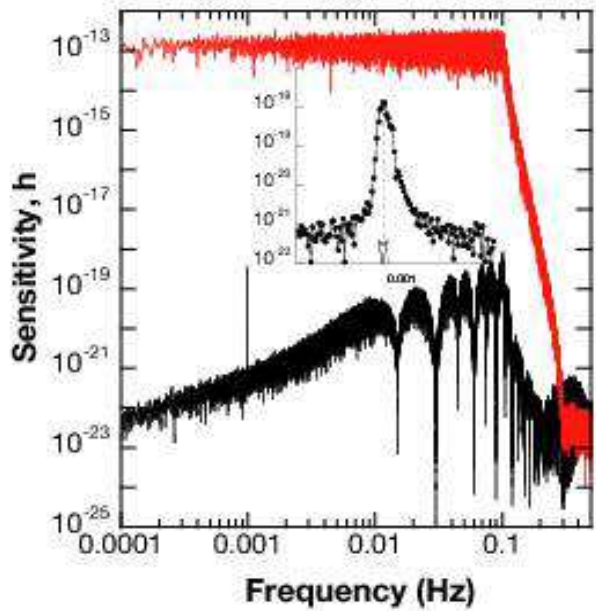} 
\caption{\small Power spectral density of the phasemeter $s_{1}$ (red curve above) and of the Michelson  
TDI X $2^{nd}$ generation (black curve below). The GW is at a frequency of $10^{-3}$ Hz and has an amplitude of $10^{-19}$. The inset shows an expansion around 1 mHz (Color on line).}
\label{PSDXGWvsNoise} 
\end{figure}

A numerical application of TDI by LISACode is illustrated in figure \ref{PSDXGWvsNoise} which shows the Power Spectral Density (PSD) of the phasemeter time-series signal (red curve above) before TDI postprocessing, and the PSD of the Michelson TDI X $2^{nd}$  generation (black curve below).  
The GW peak is clearly visible at of $10^{-3}$ Hz. The cut off observed at $0.1$ Hz is due to the  phasemeter's filter as discussed  
in Section \ref{phasemeter}. The inset in fig. \ref{PSDXGWvsNoise} shows the  structure of this peak that reflects the doppler modulation of the initial frequency due to the movements of LISA over a year. 

A way to test the numerical accuracy of LISACode is to reproduce the sensitivity curve of LISA as obtained for a given TDI combination. The sensitivity curve represents LISA's sensitivity, as a function of frequency, for an isotropic distribution of sources with random polarization. 
The sensitivity\cite{TDITinto}  corresponds to the square root of the ratio of the noise spectral density   
$PSD_{Noise}$ to the duration of the measurement (1 Year) divided by the gravitational wave TDI response $Resp^{rms}_{GW}$, and for a signal to noise ratio of 5, i.e.:  

\begin{equation} 
h = 5 \sqrt{PSD_{Noise} \over {\rm 1 Year}}\times{ 1.\over{Resp^{rms}_{GW}}}
\end{equation} 

This function is traditionally calculated using X TDI $1^{st}$ generation, for a rigid and  
non-rotating LISA, for which a semi-analytical calculation is possible  \cite{TDITinto,web,vinet_sensitivity}. This semi-analytical calculation is performed without laser noise, assuming that it is exactly cancelled by the TDI preprocessing.

As calculating with LISACode the sensitivity curve for a precise isotropic distribution of randomly polarized sources would imply a prohibitive CPU time, we have first determined a limited set of sources which yield a sensitivity curve very close to the one given by the exact calculation: an agreement better than 10\% for frequencies below $0.1 Hz$. The limited set is defined by 44 (incoherent) sources whose parameters can be obtained from\cite{LISACode_web}

\begin{figure}[!ht] 
\centering 
\includegraphics[width=8.6cm]{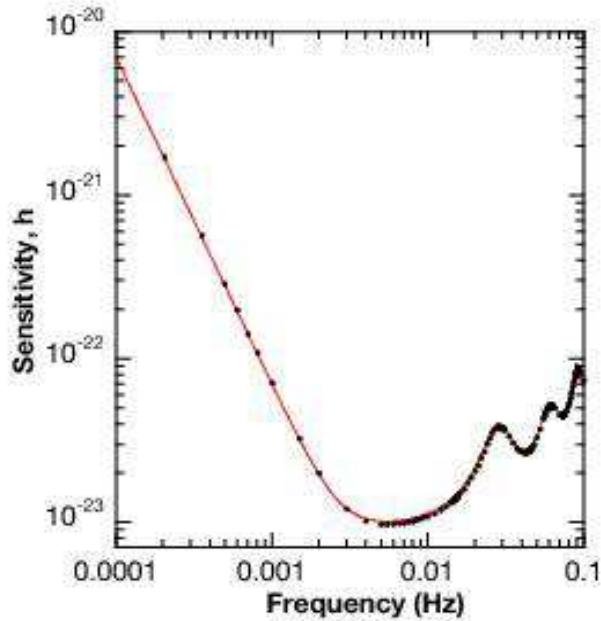} 
\caption{\small The comparison between the analytic sensitivity curve (solid red line) for an isotropic distribution and the sensitivity curve obtained with LISACode (black dots) for the same set, see section \ref{tdi}  and ref \cite{LISACode_web} (Color on line).} 
\label{SensitivityCurve_1} 
\end{figure}

\begin{figure}[!ht] 
\centering 
\includegraphics[width=8.6cm]{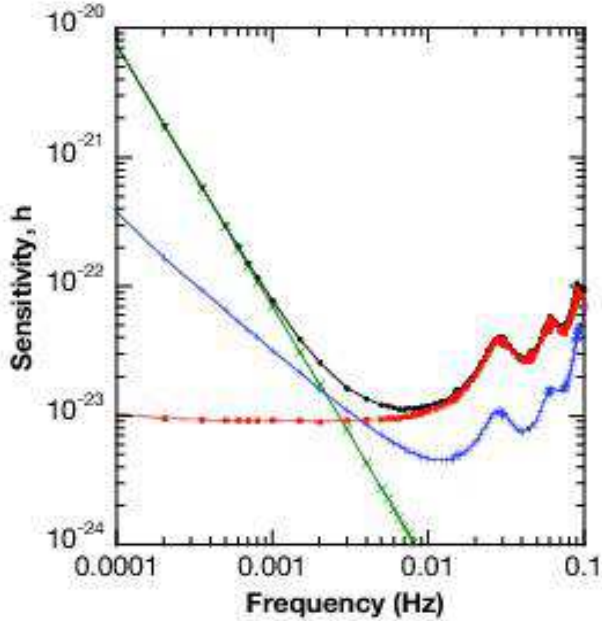} 
\caption{\small The budget of each noise contribution to the sensitivity curve of  
TDI $X$ $2^{nd}$ generation. The optical path noise including  
the shot noise are shown by (red) squares, the inertial mass noise by (green) crosses, the residual laser noise by (blue) plusses and the sum of all  by (black) dots. Curves calculated with a set of 44 sources, see section \ref{tdi}   and ref \cite{LISACode_web}(Color on line).} 
\label{NoiseImpact} 
\end{figure}

Figure \ref{SensitivityCurve_1} shows the exact semi-analytical (solid red line) calculation and the numerical calculations performed with LISACode, including laser noise, as (black) dots. As can be seen, the sensitivity obtained from the limited set reproduces very accurately the exact semi-analytical results. These calculations can be compared to results presented, for example, in \cite{SyntheticLISA} and \cite{TDITinto} (a Web \cite{web} site also allows to calculate similar sensitivity curves). These results allow us to conclude that this subset of sources can be used to test sensitivity calculations on a wide frequency range and that the implementation of TDI in LISACode properly eliminates the laser noise (see also figure \ref{NoiseImpact} for an illustration of this).  In all the following calculations, this subset of sources will be used to generate the different sensitivity curves.

\begin{figure}[!ht] 
\centering 
\includegraphics[width=8.6cm]{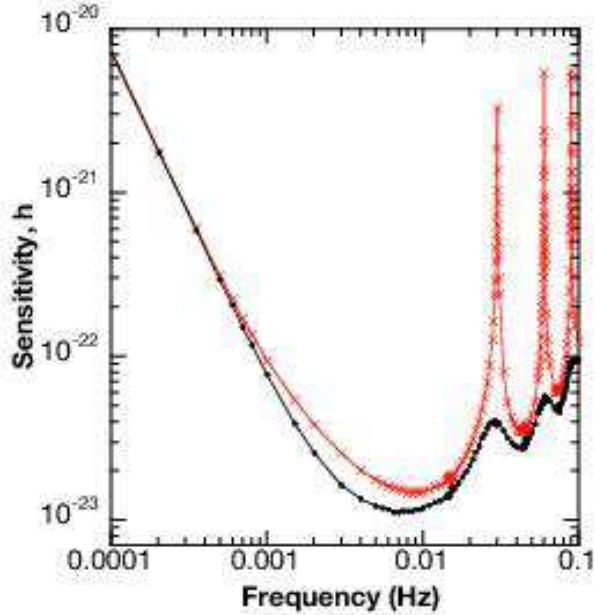} 
\caption{\small The comparison between the sensitivity curves obtained for a moving LISA. The (red) crosses show the sensitivity curve using X TDI $1^{st}$ generation and the (black) dots the sensitivity curve using  X TDI $2^{nd}$ generation. Curves calculated with a set of 44 sources, see section \ref{tdi}  and ref \cite{LISACode_web} (Color on line).} 
\label{SensitivityCurve_2} 
\end{figure}

Figure \ref{SensitivityCurve_2} compares the sensitivity curves obtained for a \emph{moving} LISA. The (red) crosses show the sensitivity curve using X TDI $1^{st}$ generation and the (black) dots the sensitivity curve using  X TDI $2^{nd}$ generation . The impact of the motion of LISA, and the imperfect annulation of the laser noise for the  $1^{st}$ generation TDI is clearly observed. The application of the  $2^{nd}$ generation combination corrects this and recovers a sensitivity curve very close to the one shown in figure \ref{SensitivityCurve_1}.

Figure~\ref{NoiseImpact} shows the noise budget of the X TDI $2^{nd}$ generation sensitivity curve.  At low frequencies, the accelerometer noise (i.e. inertial mass noise, c.f.  Table \ref{table_error}) dominates whereas at high frequencies the measurement noise, which includes the shot noise, is the most important factor. As can be seen, the application of TDI,  (assuming a laser with a power spectral density of $30 {\rm Hz/\sqrt {Hz}}$), puts the residual laser noise below the other contributions (see Section \ref{noises}). Note that this result depends on the precision with which the interpolation of the phasemeter time series is performed when using the delay operator (c.f. eq. \ref{eq_delay}). 
In order to cover the whole frequency range we have concluded, in agreement with M.Vallisneri\cite{SyntheticLISA, PostProcTDI}, that a $20^{th}$ order Lagrange interpolation is necessary.

% ************ 
% * Applications of LISACode * 
% ************ 
\section{Applications of LISACode and Pre-defined GW} 
\label{SApplicationsofLISACode} 
% 
% TDI application 
% 

The LISA community is in the process of finalizing the design of the detector and there  
are numerous areas in which a simulation software, like LISACode, can help to define the  
optimal configuration and decide between different options. In the following, we apply LISACode  
to some of these questions:  the effect of changing the nominal LISA armlength, the accuracy in the knowledge of the flightpath time on the  sensitivity of LISA and the TDI combinations that can be used in the case of a missing link.
%  
% Armlength variation 
% 
\subsection{Modifying the Armlength: from $5 \times 10^{9}$ m to $2 \times 10^{9}$  m }
\label{SSResultsArmVari} 

\begin{figure}[!ht] 
\centering 
\includegraphics[width=8.6cm]{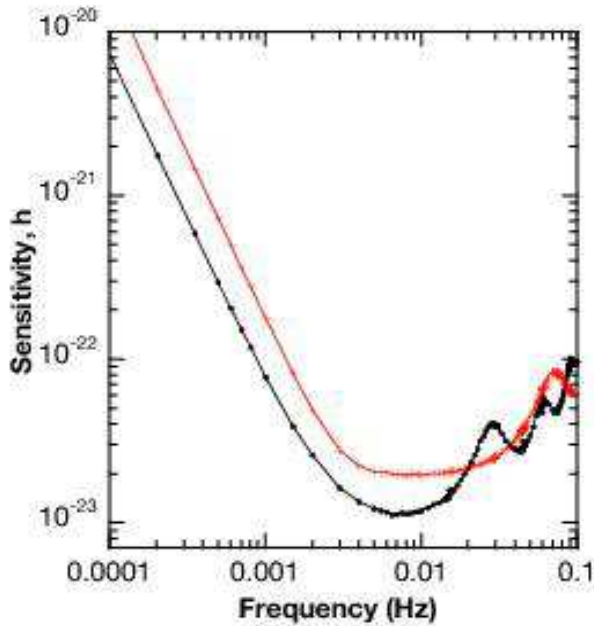} 
\caption{\small  Sensitivity curves for two armlengths: (black) dots for a $5 \times 10^{9}$ m armlength and (red) plusses for a value of $2 \times 10^{9}$ m.  Calculation are performed in the framework of TDI  $X$ $2^{nd}$ generation. Curves calculated with a set of 44 sources, see section \ref{tdi}  and ref \cite{LISACode_web}(Color on line).} 
\label{Sensitivity_VarL} 
\end{figure}

The scale of LISA (nominal armlength) has been a subject of debate within the LISA  
community as it has sometimes been advocated that a $smaller$ LISA could result in a  
reduction of its financial budget. LISACode can address the scientific aspect of this  
question which is relative to the dependence of the sensitivity curve on the nominal  
armlength. Figure \ref{Sensitivity_VarL} compares the sensitivity curves for armlengths  
of $2\times10^{9}$ m  and $5\times10^{9}$ m. One observes that the sensitivity for the shorter armlength is shifted towards higher frequencies and that the ``oscillation'' peaks, conditioned by the armlength value, are positioned at the corresponding frequencies. 

% 
% Inaccuracy on the delays 
% 
\subsection{Inaccuracy on the delays} 
\label{SSResultsInaccuracyDelays} 
\begin{figure}[!ht] 
\centering 
\includegraphics[width=8.6cm]{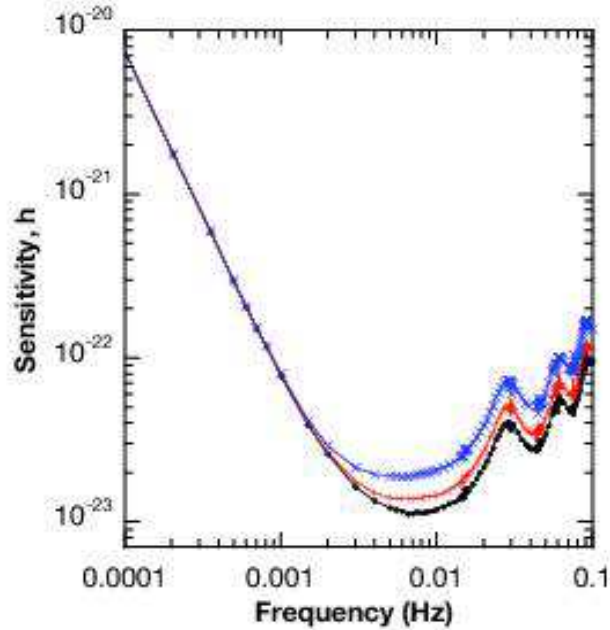} 
\caption{\small The sensitivity curves for different offsets on the laser travel time used in TDI: 0 (black dots), 0.5 (red plusses) and 1 $\mu sec$ (blue crosses). Calculations are performed for TDI  $X$ $ 2^{nd}$ generation. Curves calculated with a set of 44 sources, see section \ref{tdi}  and ref \cite{LISACode_web} (Color on line).} 
\label{Sensitivity_DeltaL} 
\end{figure}

Application of TDI requires a very good precision on the travel time of the lasers beams 
between spacecraft (see Equation \ref{TDIGeneratorX} and the discussion in \cite{TDIVinet2}).  
This travel time will be known with a given precision and the sensitivity of LISA will depend  
on this precision. In order to test the sensitivity as a function of the precision, an offset  
has been added to the exact delay used in a given TDI combination: 
\begin{equation} 
D_{TDI} = D_{exact} + \Delta D_{offset} 
\end{equation} 
This offset, $\Delta D_{offset}$, is assumed to have a magnitude of about $1  \; \mu$sec,  
i.e. a photon flight path of about $300$ m.  
Figure~\ref{Sensitivity_DeltaL} shows the evolution of the sensitivity for  
TDI $X$ $2^{nd}$ generation for an offset of $0$, $0.5$ and $1 \; \mu$sec respectively.  
As the level of the laser noise is of the same order of magnitude as the level of the other noises,  
the sensitivity curve goes up as a function of the offset due to the increasing error  
in the interpolation scheme produced by the inaccuracy of the delay.

%  
% Missing a link 
% 
\subsection{The effect of a missing link}
\label{SSMissingLink} 

\begin{figure}[!ht] 
\centering 
\includegraphics[width=8.6cm]{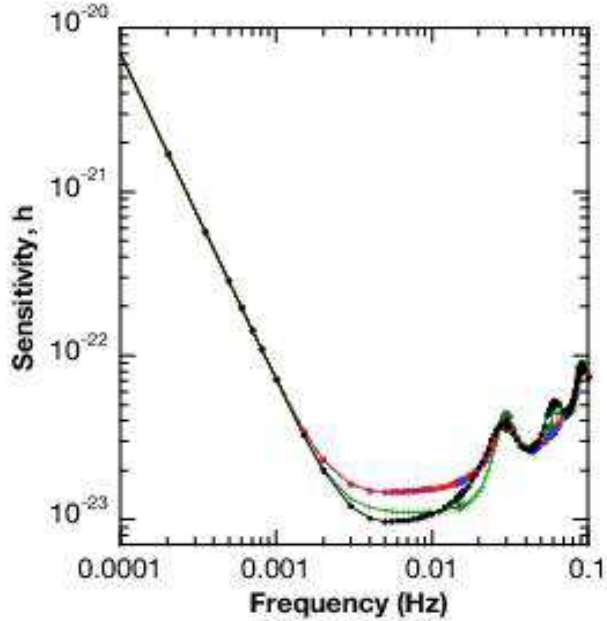} 
\caption{\small The comparison of sensitivity curves for TDI combinations  
Beacon (red squares), Monitor (blue crosses) and Relay (green plusses) with  X $1^{st}$ (black dots). Curves calculated with a set of 44 sources, see section \ref{tdi}  and ref \cite{LISACode_web} (color on line)} 
\label{BeaconRelayMonitor} 
\end{figure}

A question that the LISA mission may face is that of the loss of a $link$ (from spacecraft $i$ to $j$, but still keeping the link of $j$ to $i$) between spacecraft since the question of the reliability of the links has also an important impact on its financial budget. Several TDI combinations have the property of allowing for the loss of such a link. Figure \ref{BeaconRelayMonitor} shows the Beacon, Monitor and Relay combinations\cite{TDIMoving} and the Michelson $1^{st}$ generation which requires all the links between one spacecraft and the two other. If all the links are available, all three Michelson ($X,Y$ and $Z$) would be conserved whereas, in the case of the loss of a link , one would conserve only one Michelson and the Beacon, Monitor and Relay combinations (Note that the Beacon and Monitor are numerically almost indistinguishable). The loss of precision in determining the physical parameters of a source when a link is loss has still to be determined.

%
%  Pre-defined GW
%

\section{Pre-defined GW in LISACode} 
\label{SSSPredefinedGW} 

The LISACode software package includes a number of predefined types of gravitational waves  
which are rapidly described below. 
\subsection{A monochromatic wave:} 
This type of wave is essentially a ``mathematical'' object which is directly defined in the  
Canonical Reference Frame (see fig. \ref{GWParameters}). Therefore the inputs to LISACode  
will be  the amplitudes $h_{0+}$ and $h_{0\times}$ of the two polarization components,  
 the frequency $f$ and  $\phi_{0+}$ and $\phi_{0\times}$ the initial phases of the two  
polarization components. The conventions for the strain calculation are: 
\begin{equation} 
\left\{ \begin{array}{lll}  
{h}_{CRF+} (t) & = & h_{0+}  \sin \left( 2 \pi f t + \phi_{0+}\right)\\ 
h_{CRF\times} (t) & = & h_{0\times}  \sin \left( 2 \pi f t + \phi_{0\times}\right) 
\end{array} \right.  
\label{GWMono} 
\end{equation}\\ 

\subsection{GW signal from a binary with fixed frequency: } 
A binary of fixed frequency goes one step further. LISACode defines the amplitude by  
the physical parameters of the system in the canonical reference frame 
(see Appendix \ref{SSSFrames} and fig. \ref{GWParameters}):  
\begin{equation} 
\left\{ \begin{array}{lll}  
h_{CRF+} &=& A \left(1 + \cos^2 i \right) \cos \left( 4 \pi f_{orb} t + \phi_0 \right)\\ 
h_{CRF\times} &=& 2 A \cos i  \sin \left( 4 \pi f_{orb} t + \phi_0 \right) 
\end{array} \right.  
\label{GWBinaryFixedFreq} 
\end{equation} 
with 
\begin{eqnarray} 
m_{tot} &=& m_1 + m_2 \label{mtot}\\ 
R &=& {\left( { G m_{tot} \over {\left( 2 \pi f_{orb} \right)}^2} \right)}^{1/3} \label{RBinFixF}\\ 
A &=& {2 G^2 \over c^4} {m_1 m_2 \over R r} \label{ABinFixF} 
\end{eqnarray} 
where $m_1$ and $m_2$ are the masses of the two objects, $f_{orb}$ is the orbital frequency,   
$i$ the inclination, $\phi_{0}$ the initial phase and $r$ the distance from the source to LISA 
($c$ is the light velocity). 

\subsection{GW signal from a coalescent binary, computed in the Post-Newtonian approximation: } 
\begin{figure}[!ht] 
\centering \includegraphics[angle=270, width=8.6cm]{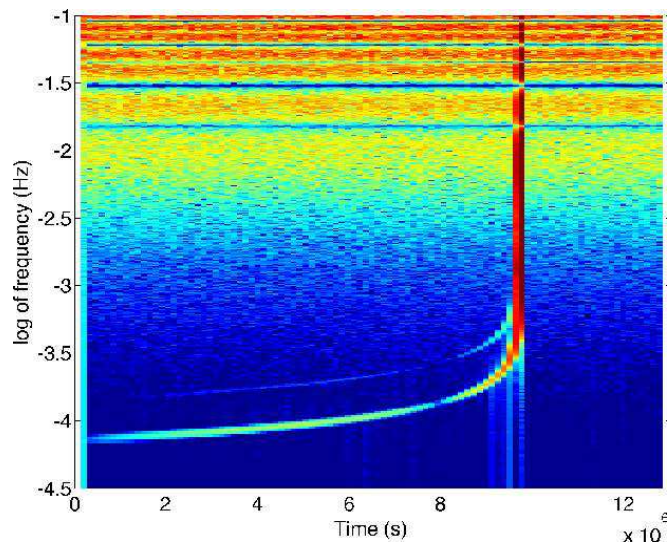} 
\caption{\small Time frequency figure of the Michelson TDI ($X$ $2^{nd}$ generation) for a binary system computed in Post-Newtonian approximation at 2.5 PN during 5 months ($10^{6}$ - $10^{6}$ solar masses, coalescence time 4 months, inclination 90 degrees and distance 100 Mpc). The color scale represents the log of the amplitude (Color on line).} 
\label{GWPN25} 
\end{figure}

The input parameters are the same as above except for the additional parameter  
$T_{coalescence}$ which defines the coalescence time.  
Figure \ref{GWPN25}  shows an example (time-frequency matrix for a 5 month period) of such a calculation superposed on the LISA instrumental noise. Reference \cite{2PN} gives an explanation of the model which is used.  

\subsection{Input of a GW from a Text file.} 
\label{SSSGWFile} 
In the pre-defined GWs described above, the direction of the angular momentum of  
the emitting system does not depend on time and, therefore, neither do the unit  
vectors $ \widehat{p}$ and $ \widehat{q}$ and, hence, the inclination and  
polarization angles.

\begin{figure}[!ht] 
\centering \includegraphics[angle=270, width=8.6cm]{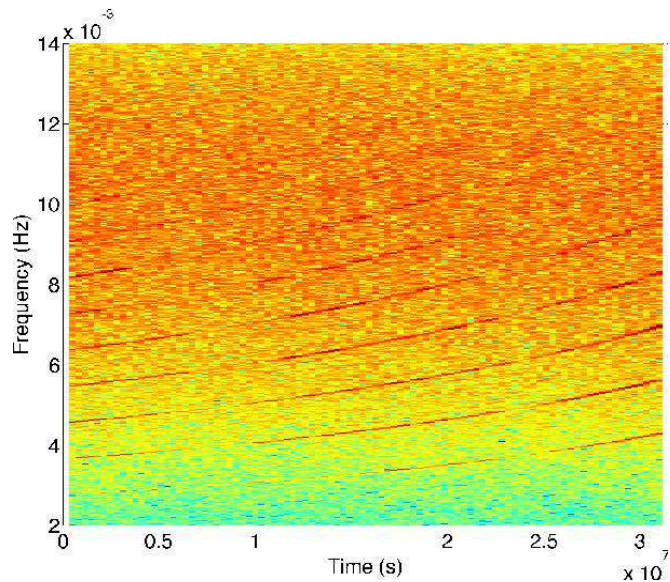} 
\caption{\small Time frequency figure of the Michelson TDI ($X$ $2^{nd}$ generation) for EMRI during 1 year ($10$ - $10^{6}$ solar masses, black hole spin ratio 0.8 and distance 100 Mpc ). The color scale represents the log of the amplitude (Color on line).} 
\label{TFEMRIs} 
\end{figure}

In more complicated cases, such as for EMRIs (Extreme Mass Ratio Inspiral) \cite{BarackCutler} ,  
this is not the case and therefore besides $h_{CRF,+}$ and $h_{CRF,\times}$,  
the inclination and polarization angles would have to be given as a function of time.  
To simplify the inputs, LISACode requests that the strain be expressed directly in the  
``Observational Reference Frame'' (ORF, see fig. \ref{GWParameters}).  
The requested (ASCII) input file should therefore contain, for each time step, a line  
giving the time, $h_{ORF+}(t)$ and $h_{ORF\times}(t)$. The data can be given with any  
constant time step as LISACode will linearly interpolate this data to obtain data sampled at   
the correct, user-defined, phasemeter time step. More details on the input file can be  
found in \cite{LISACode_web}. Figure \ref{TFEMRIs} shows an example (Time-Frequency matrix for a one year period, including the detector noise) of harmonics produced by an EMRI with a mass ratio of $10^5$, a spin ratio of $S/M=0.8$, at a distance of 100 Mpc. 

% *************** 
% * Conclusion * 
% *************** 
\section{Conclusion and perspectives} 
\label{SConclusion} 

We have presented a new simulator (LISACode) whose ambition is to achieve, through 
a very modular structure, a new degree of sophistication allowing to map, as closely 
as possible, the precise impact of the different sub-systems on the precision of 
LISA measurements. LISACode is considered as a tool whose purpose is to bridge the gap between the basic principles 
of LISA and a, future, sophisticated end-to-end simulator. This is achieved by 
introducing, in a realistic manner, most of the ingredients that will influence 
its sensitivity. Most of the parameters of a LISA configuration (orbits, noise functions,...) 
can be user-defined. LISACode can use number of pre-defined gravitational waveforms  
but more complicated waveforms can also be introduced via text files. As the suppression 
of the laser noise is a very important aspect of LISA, LISACode allows to implement 
a number of pre-defined TDI combinations and allows the user to construct more 
specific combinations if needed. In its most complete form, LISACode is relatively 
time consuming but different  and faster versions of LISACode can be constructed 
and used, for example, for data analysis (see Appendix \ref{CPU}).

% ************ 
% * Reference Frames and Gravitational Waves * 
% ************ 

\appendix

\section{Reference frames and the response of LISA to a GW} 
\label{SSSFrames} 
There are a number of reference frames  which have to be defined  in order to properly  
understand and generate the response of LISA (see for instance :  
\cite{Appostolatos,SyntheticLISA,LISASimulator,TDIVinet}).

The first one is the \textbf{Barycentric Ecliptic Frame} (BEF), centered on the Sun.  
Its X-axis is defined by the direction pointing towards the vernal point.  The Z-axis is defined by the direction of the ecliptic pole and the Y-axis completes the reference frame.

\begin{figure}[!ht] 
\centering \includegraphics[width=8.6cm]{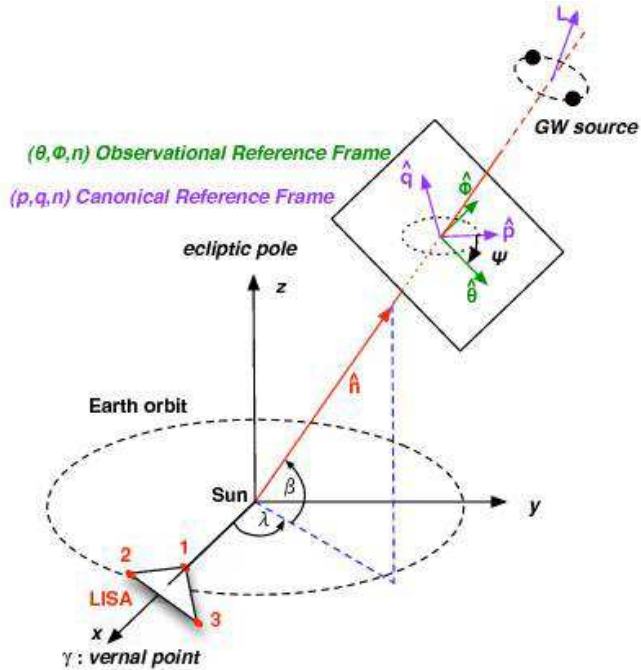} 
\caption{\small Schematic description of GW parameters and reference frames.  
The direction of the source is located by the declination $\beta$ and the ecliptic  
longitude $\lambda$. ($ \widehat{\theta}$, $ \widehat{\phi}$, $\widehat{n}$)  define the  
Observational Reference Frame , i.e. the frame  constructed from the direction of the source.  
In this frame $ \widehat{\theta}$ is on the meridian.   
($ \widehat{p}$, $ \widehat{q}$, $\widehat{n}$) define the Canonical Reference Frame.  
$\psi$ is the polarization angle, i. e. the angle between the Canonical Reference Frame   
and the Observational Reference Frame (Color on line).} 
\label{GWParameters} 
\end{figure}

In this frame, the source direction $\widehat{n}$ is defined  by the declination $\beta$  
and the longitude $\lambda$, as shown in fig. \ref{GWParameters}.\emph{ This is the reference frame that is used by LISACode for computation.}

Using this direction, an orthonormal frame  
($ \widehat{\theta}$, $ \widehat{\phi}$, $\widehat{n}$),  
called the \textbf{Observational Reference Frame} (ORF) is defined by: 
\begin{equation} 
\widehat{n}  
= \begin{pmatrix} 
\cos \beta \cos \lambda\\  
\cos \beta  \sin \lambda\\  
\sin \beta \end{pmatrix} 
,\; 
\widehat{\theta}  
= -{\partial \widehat{n}\over \partial \beta} 
= -\begin{pmatrix} 
\sin \beta \cos \lambda\\  
\sin \beta  \sin \lambda\\  
- \cos \beta 
 \end{pmatrix} 
,\; 
\widehat{\phi} 
= {-1 \over \cos \beta} {\partial \widehat{n}\over \partial \lambda} 
= \begin{pmatrix} 
 \sin \lambda\\ 
- \cos \lambda\\  
0 
\end{pmatrix} 
\label{BaseOnDirection} 
\end{equation} 
The $ \widehat{\theta}$, $ \widehat{\phi}$ plane is called the observation plane.\
Also defined by the source direction, the \textbf{Canonical Reference Frame} (CRF: $ \widehat{p}$, $ \widehat{q}$, $\widehat{n}$) is defined. The unit vector $\widehat{p}$ is defined as the  direction of the major axis of the ellipse obtained by the projection of the orbit on the observation plane. The definition of $\widehat{q}$ follows from the product  $\widehat{p} \times\widehat{n}$.

Geometrically, and in the case of a circular orbit, the unit vectors  
$ \widehat{p}$ and $ \widehat{q}$ (labelled  $ \widehat{a}$ and $ \widehat{b}$  
in \cite{TDIVinet}) are aligned with the major and minor axis of the projection  
of the orbit on the $ \widehat{\theta}$, $ \widehat{\phi}$ plan. In terms of polarization,  
$ \widehat{p}$ and $ \widehat{q}$ define the reference frame within which the gravitational  
strain is given by: 

\begin{equation} 
h_{ij} (t) 
= \begin{pmatrix} 
h_{CRF+}(t) & h_{CRF\times}(t) & 0 \\ 
h_{CRF\times}(t) & -h_{CRF+}(t) & 0 \\ 
0 & 0 & 0  
\end{pmatrix} 
\label{GWCompInFrame} 
\end{equation} 
Where $h_{CRF+}(t)$ and  $h_{CRF\times}(t)$  are the time evolution of the plus and  
cross polarization components in the canonical frame (CRF). 

The inclination angle $i$ is defined as the angle between $\widehat{L}$  
(orbital angular momentum of the emitting system) and the source direction $\widehat{n}$. 

The polarization angle, $\psi$, is defined as the angle between  $\widehat{\theta}$  
and  $\widehat{p}$, i.e.: 
\begin{equation} 
\left\{ \begin{array}{l}  
\widehat{p} = \cos \psi \; \widehat{\theta} - \sin \psi \; \widehat{\phi}  \\  
\widehat{q} = \sin \psi \; \widehat{\theta} + \cos \psi \; \widehat{\phi}  
 \end{array} \right. 
\label{PsiPassRef} 
\end{equation} 

It is important to notice that in many cases, $ \widehat{p}$, $ \widehat{q}$,  
the polarization angle and the inclination angle will be time dependent and will  
therefore imply special care in defining the gravitational wave strain for LISACode.  
An example of this is given in section \ref{SSSGWFile} for an EMRI where a GW  
is loaded to LISACode via a text file. 

% ************ 
% * Response to a Gravitational Waves * 
% ************ 

The strain of a GW produces a phase fluctuation in the laser link between two spacecraft.  
Since this phase fluctuation is related to the frequency fluctuation, the LISA response involves 
 the six frequency fluctuations of the laser links. The relative GW frequency  
fluctuation for a laser received by  spacecraft B from  spacecraft A, is directly  
connected to the gravitational strain  \cite{TDIVinet} by~: 
\begin{equation} 
\left.{\delta \nu \over {\nu}_{opt}}\right|_{AB} (t) = { H_{AB}\left(t+ \widehat{n}  \cdot 
\overrightarrow{r_{B}} \right) - H_{AB}\left(t+ \widehat{n}  \cdot 
\overrightarrow{r_{A}} - {L_{AB} / c}\right) \over 2 \left(1 +  \widehat{n} \cdot  
\widehat{n_{AB}} \right)} 
\label{RepDnuonu} 
\end{equation} 
with  
\begin{equation} 
H_{AB}(t) = {h}_{ORF+}(t) \; \xi_{+,AB} + {h}_{ORF\times} (t) \; \xi_{\times,AB} 
\label{Hfctxi} 
\end{equation} 
where $ \widehat{n}$ is the unit vector of the source's direction,  
$\overrightarrow{r_{A}}$ and $\overrightarrow{r_{B}}$ are the position of the  
spacecraft in the barycentric reference frame, $L_{AB} / c$ is the time of propagation of a photon between the two spacecraft, $\widehat{n_{AB}}$ is the unit vector along arm AB, ${h}_{ORF+}(t)$  
and ${h}_{ORF\times}(t)$ the time evolutions of the GW's two components in the  
observational frame ($ \widehat{\theta}$, $ \widehat{\phi}$, $ \widehat{n}$). $\xi_{+,AB}$  
and $\xi_{\times,AB}$ are the two antenna pattern functions:   
\begin{equation} 
\left\{ \begin{array}{l}  
\xi_{+ i} = {\left(\widehat{\beta}  \cdot \widehat{n_i} \right)}^2 - {\left(\widehat{\lambda} \cdot\widehat{n_i} \right)}^2\\  
\xi_{\times i } = 2 \left(\widehat{\beta} \cdot\widehat{n_i} \right) \left(\widehat{\lambda} \cdot\widehat{n_i} \right) 
\end{array} 
\right. 
\label{} 
\end{equation}\\ 
The relation between the GW's two components in the observational frame (ORF: $ \widehat{\theta}$, $ \widehat{\phi}$, $ \widehat{n}$) and the components in the canonical frame (CRF: $ \widehat{p}$, $ \widehat{q}$, $ \widehat{n}$) is: 
\begin{equation} 
\left\{ \begin{array}{lll}  
{h}_{ORF+} & = & \cos({2\psi}) \; h_{CRF+} +  \sin({2\psi}) \; h_{CRF\times}\\ 
{h}_{ORF\times} & = & -\sin({2\psi}) \; h_{CRF+} +  \cos({2\psi}) \; h_{CRF\times} 
\end{array} \right.  
\end{equation} 

\section{CPU considerations} 
\label{CPU}
In its complete configuration, LISACode aims at  generating  
time sequences of data with $realistic$ noise functions  and  testing the impact  
of different LISA  
configurations: orbits, TDI combinations, etc... 

It soon appeared necessary to use LISACode within a data analysis strategy. For this, the CPU  
execution time is a critical issue sometimes overriding the priority of numerical precision.  
For instance, assuming exact cancellation of laser noises saves a significant amount of  
CPU time.  
Other  
options as interpolation of the orbits and using less efficient but simpler 
 TDI combinations  
can also  
speed up the execution. Table \ref{table_speed} gives some execution times, measured on a G5  
biprocessor Mac ($2.7$ GHz) running Unix. The last line  corresponds to using a simplified  
version, LISALight, which is compiled with modules extracted from the LISACode libraries in order to directly calculate the TDI combinations.  Calculations are performed for a monochromatic wave of frequency $1 mHz$.

\begin{table}[htdp]
\caption{Execution time for LISACode simulations for one year with a time sampling of $\Delta  t_{sampling}$. Order 2 orbits take into account flexing, Sagnac, aberration and relativistic gravitational effects. Order 0 orbits include flexing only. LISALight corresponds to a program compiled from the LISACode libraries but simplified as much as possible. Calculation were performed on a G5 biprocessor Mac ($2.7\; GHz, 5\; Go\;RAM $) running Unix.}
\begin{center}
\begin{tabular}{|c|c|c|c|c|c|c|}
\hline
& $\Delta  t_{Sampling}$ & Orbits & Noises & Interpolation & TDI &  CPU time  \\
& (seconds) & (order) &   &(order)   &   &  (seconds)  \\
\hline
1&1 & 2 & All & 20 & $X^{2nd}, Y^{2nd}, Z^{2nd}$ & 15Ê299  \\
2&1  & 2 & All & 8 & $X^{2nd}, Y^{2nd}, Z^{2nd} $& 8Ê143 \\
3&1  & 2 & All & 8 & $X^{2nd} $& 4Ê652 \\
4&1  & 2 & no laser noise & 8 &$ X^{2nd} $& 3Ê891  \\
5&15  & 2 & no laser noise & 8 &$ X^{2nd} $  & 242  \\
6&15  & 0& no laser noise & 8 &$ X^{1st} $& 140  \\
7&15  & 0& none& 2 &$ X^{1st} $& 74 \\
8&120  & 0& none& 2 &$ X^{1st} $& 7.8 \\
\hline
 &  &  & LISALight & && \\
\hline
9&120  & 0 & none & none & $X^{1st}$ & 2.3 \\
%11&120  & 0 & none & none & $Michelson$ & 1.8 \\
\hline
\end{tabular}
\end{center}
%\label{default}
\label{table_speed}
\end{table}%

\bibliographystyle{plain}

\end{document}